\documentclass[final,3p,times,twocolumn]{elsarticle}
\usepackage{graphicx}
\usepackage{amsmath}
\usepackage{amssymb}
\usepackage{hyperref}

\begin{document}

\title{Critical behavior in itinerant ferromagnet SrRu$_{1-x}$Ti$_x$O$_3$}

\author{Renu Gupta}
\author{Imtiaz Noor Bhatti}
\author{A. K. Pramanik\corref{cor1}}
\ead{akpramanik@mail.jnu.ac.in}

\address{School of Physical Sciences, Jawaharlal Nehru University, New Delhi - 110067, India.}

\begin{abstract}
SrRuO$_3$ presents a rare example of ferromagnetism among the 4$d$ based oxides. While the nature of magnetic state in SrRuO$_3$ is mostly believed to be of itinerant type, recent studies suggest a coexistence of both itinerant and localized model of magnetism in this material. Here, we have investigated the evolution of magnetic state in doped SrRu$_{1-x}$Ti$_x$O$_3$ through studying the critical behavior using standard techniques such as, modified Arrott plot, Kouvel-Fisher plot and critical isotherm analysis across the magnetic transition temperature $T_c$. The substitution of nonmagnetic Ti$^{4+}$ (3$d^{0}$) for Ru$^{4+}$ (4$d^4$) would simply dilute the magnetic system apart from modifying the electron correlation effect and the density of states at Fermi level. Surprisingly, $T_c$ does not change with $x$. Moreover, our analysis show the exponent $\beta$ related to spontaneous magnetization increases while the exponents $\gamma$ and $\delta$ related to initial inverse susceptibility and critical magnetization, respectively decrease with Ti substitution. The estimated exponents do not match with any established theoretical models for universality classes, however, the exponent obey the Widom relation and the scaling behavior. Interestingly, this particular evolution of exponents in present series has similarity with that in isoelectronic doped Sr$_{1-x}$Ca$_x$RuO$_3$. We believe that site dilution by Ti leads to formation magnetic clusters which causes this specific changes in critical exponents.
\end{abstract}


\maketitle
\section {Introduction}
The 4$d$ based perovskite SrRuO$_3$ continues to attract large deal of scientific attention which includes both fundamental as well as technological interest. This material is commonly believed to be an itinerant type ferromagnet with transition temperature around 160 K, while offering a rare example of 4$d$ based oxide having ferromagnetic (FM) ordering.\cite{allen,cao,fuchs,mazin,kim,gupta,cheng} The itinerant nature of magnetic state is manifested in the fact that measured moment shows a lower value, $\sim$ 1.4 $\mu_B$/f.u. in magnetic field as high as 30 Tesla compared to expected spin-only value, 2 $\mu_B$/f.u. for $S$ = 1.\cite{cao} Interestingly, a recent theoretical study\cite{kim} has predicted a coexistence both itinerant and localized nature of magnetism in SrRuO$_3$ which has also been experimentally discussed in our previous study.\cite{gupta} Furthermore, debate continues about the nature of magnetism in SrRuO$_3$. While the majority of studies report mean-field like magnetic state in SrRuO$_3$,\cite{cheng,fuchs,kim-crit} there are several studies which imply 3D Heisenberg- or Ising-type spin interaction in this material.\cite{palai,kats,klein,chanchal} Interestingly, one recent study shows that linearity in Arrott plot (signature for mean-field model) in SrRuO$_3$ is mainly realized due to continuous curvature evolution from Ca$_{0.5}$Sr$_{0.5}$RuO$_3$ to BaRuO$_3$ via SrRuO$_3$ as lattice distortions and spin-orbit coupling changes, and it is less likely due to itinerant type ferromagnetism in SrRuO$_3$.\cite{cheng} Similarly, influence of anisotropy on critical behavior where the exponent values change with crystal axis has also been shown for SrRuO$_3$.\cite{klein,wang} This underlines the fact that even after large volume of study, the detail nature of magnetic state in SrRuO$_3$ is still debated.

There have been several attempts to understand the nature of magnetism using route of chemical substitution. The most prominent one is the isoelectronic substitution at Sr-site. The Ca$^{2+}$ substitution in Sr$_{1-x}$Ca$_x$RuO$_3$ shows a total suppression of FM ordering at about 70\% of doping concentration, a phenomenon which has been associated with the FM quantum phase transition (QPT) phenomenon.\cite{fuchs,cao} Band structure calculations show that substitution of Ca causes further distortion in RuO$_6$ octahedra and in Ru-O-Ru bond angel which effectively decreases the density of states at Fermi level N($\epsilon_F$), hence the necessary Stoner criterion for itinerant ferromagnet is no more satisfied.\cite{mazin} The critical behavior in Sr$_{1-x}$Ca$_x$RuO$_3$ shows an interesting evolution where the exponent $\beta$ increases, and both $\gamma$ and $\delta$ decreases with Ca doping which has been attributed to effect of disorder arising from quantum fluctuation close to QPT point and phase segregation effect.\cite{fuchs,cheng} On other hand, Ba$^{2+}$ substitution in Sr$_{1-x}$Ba$_x$RuO$_3$ lowers the $T_c$ down to about 60 K and the nature of magnetism is found to closely follow the 3D Heisenberg model.\cite{jin}

In present work, we have investigated the magnetic state in SrRu$_{1-x}$Ti$_x$O$_3$ by studying an evolution of critical behavior as the related critical exponents and critical temperature represent an intrinsic nature of magnetic behavior of a material. From structure wise, ionic radii of Ru$^{4+}$ (0.62 \AA) and Ti$^{4+}$ (0.605 \AA) closely match which implies this substitution will introduce minimum structural modification, hence the structural disorder induced modification in magnetic state is least expected. Rather, nonmagnetic Ti$^{4+}$ would simply dilute the magnetic structure formed by transition metal and oxygen network. Further, substitution of Ti$^{4+}$ (3$d^0$) for Ru$^{4+}$ (4$d^4$) would oppositely tune the electron correlation $U$ and N($\epsilon_F$) which will have wide ramification on Stoner criteria of itinerant FM i.e., $U$N($\epsilon_F$) $>$ 1.\cite{stoner} Even, a large change of $T_c$ has been observed with variation of N($\epsilon_F$) in ultra thin film of SrRuO$_3$.\cite{chang} In fact, recently we have shown while effective magnetic moment and the Curie-Weiss temperature decreases in SrRu$_{1-x}$Ti$_x$O$_3$ with $x$, the FM transition temperature $T_c$ appears to remain unchanged.\cite{gupta} This unchanged behavior of $T_c$ has been understood through an opposite tuning of $U$ and N($\epsilon_F$) in picture of itinerant ferromagnet where the combined term $U$N($\epsilon_F$) effectively remains constant with Ti doping. In deed, photoemission spectroscopy measurements as well as band structure calculations have shown gradual increase of $U$ and depletion of N($\epsilon_F$) with Ti substitution in SrRuO$_3$.\cite{kim1,lin} With Ti substitution, SrRuO$_3$ further develops Griffiths-phase like behavior which arises due to disorder coming from formation of magnetic clusters above $T_c$ which has similarly been evidenced in Sr$_{1-x}$Ca$_x$RuO$_3$.\cite{jin} This present series of samples share some of the properties with isoelectronic doped Sr$_{1-x}$Ca$_x$RuO$_3$, therefore it would be interesting to understand the critical behavior in SrRu$_{1-x}$Ti$_x$O$_3$. 
  
Here, we have studied the critical behavior in SrRu$_{1-x}$Ti$_x$O$_3$ series with $x$ = 0.0, 0.1, 0.3, 0.4, 0.5 and 0.7. We have estimated the critical exponents ($\beta$, $\gamma$ and $\delta$) and $T_c$ following various independent methods such as, modified Arrott plot, Kouvel-Fisher method and critical isotherm analysis. The estimated exponent $\beta$ for SrRuO$_3$ is close to the value for mean-field model (Table I). The exponents for doped materials do not match with the values theoretically predicted for different universality classes based on 3-dimensional magnetism. The estimated exponents, however, obey the scaling law behavior and Widom relation which implies values are correct.

\section {Experimental Details}
Polycrystalline samples of series SrRu$_{1-x}$Ti$_x$O$_3$ with $x$ = 0.0, 0.1, 0.3, 0.4, 0.5 and 0.7 are prepared by standard solid state method. The samples have been characterized by x-ray diffraction (XRD) and by Rietveld analysis of XRD data. All the samples are in single phase and without any noticeable chemical impurity. Details of sample preparation and characterization are given elsewhere. \cite{gupta} Temperature ($T$) dependent magnetization ($M$) data have been collected with superconducting quantum interference device (SQUID) magnetometer by M/s Quantum Design. For critical analysis,  magnetic field ($H$) dependent isotherms $M(H)$ have been collected at an interval of 1 K across $T_c$ using vibrating sample magnetometer (VSM) by M/s Cryogenics Ltd. For proper stabilization of temperature, about 10 minute wait time has been given before recording each isotherm. The external applied magnetic field ($H_a$) has been corrected by the demagnetization effect to get the internal magnetic field $H_i$ [=$H_a$-N M(T,$H_a$)], where M(T,$H_a$) is the measured magnetization and N in the demagnetization constant that has been calculated from physical dimensions of sample. \cite{osborn} This calculated $H_i$ has been used for critical exponent scaling analysis. The critical temperature and critical exponents are have been determined by commonly used techniques like modified Arrott plots (MAP), Kouvel-Fisher (KF) method and critical isotherm analysis.

\section{Results and Discussions}
\subsection{Scaling Analysis}
In case of second-order phase transition, the correlation length ($\xi$) among spins diverges at the magnetic phase transition temperature $T_c$ following $\xi$ = $\xi_0|1 - (T/T_c)|^{-\nu}$ where $\nu$ is the exponent. Following this, scaling hypothesis predicts that spontaneous magnetization $M_s$ below $T_c$, initial inverse susceptibility $\chi_0^{-1}$ above $T_c$ and magnetization at $T_c$ obey set of power law behavior with temperature as described below, \cite{stanley}   

\begin{eqnarray}
	M_s(T) = M_0(-t)^\beta,  t < 0
\end{eqnarray}   

\begin{eqnarray}
	\chi_0^{-1}(T) = G(t)^\gamma,  t > 0
\end{eqnarray} 

\begin{eqnarray}
	M = XH^{1/\delta},  t = 0
\end{eqnarray}

where $t$ = (T - T$_c$)/T$_c$ is the reduced temperature; $M_0$, $G$ and $X$ are the critical amplitudes and $\beta$, $\gamma$ and $\delta$ are the critical exponents. The scaling hypothesis further predicts magnetic equation of state which describes the relationship between $M(H,t)$, $H$ and $T$ in following mannar, 
			
\begin{eqnarray}
	M(H,t) = t^{\beta} f_\pm\left(\frac{H}{t^{\beta+\gamma}}\right) 
\end{eqnarray}	
	
where $f_+$ and $f_-$ are the regular functions for $T > T_c$ and $T < T_c$, respectively. The Eq. 4 implies that for right values of critical temperature $T_c$ and critical exponents $\beta$ and $\gamma$, the scaled magnetization $m$ = $t^{-\beta}$$M(H,t)$ plotted as a function of scaled field $h$ = $t^{-(\beta+\gamma)}$$H$ would fall on two distinct curves for isotherms both above and below $T_c$. 

\subsection{DC Magnetization study}
Fig. 1 shows dc magnetization data collected following field cooled (FC) protocol in magnetic field of 100 Oe for SrRu$_{1-x}$Ti$_x$O$_3$ series. Recently, we have shown an interesting evolution of magnetic behavior of SrRu$_{1-x}$Ti$_x$O$_3$ series.\cite{gupta} For parent SrRuO$_3$, we have observed the magnetic transition temperature $T_c$ $\sim$ 163 K which is in conformity with other studies.\cite{allen,cao,fuchs,chanchal} With site dilution through Ti substitution, while though the Curie-Weiss temperature ($\theta_P$) and the magnetic moments decreases, interestingly the long-range magnetic ordering temperature $T_c$ does not change. In scenario of itinerant ferromagnetism, we explained this constant nature of $T_c$ through opposite tuning of density of states (DOS) and electronic correlation ($U$) with Ti substitution. As evident in main panel of Fig. 1, FC magnetization for all the samples show a sudden rise around 163 K which marks the $T_c$. The inset of Fig. 1 depicts ZFC magnetization data in a close temperature scale around $T_c$. As evident in inset figure, $M_{ZFC}$ show peak around 163 K which does not change with $x$. Here, it can be mentioned that our critical analysis of $M_{FC}(T)$ data also shows a constant $T_c$ in present SrRu$_{1-x}$Ti$_x$O$_3$ series. This analysis further shows critical exponent $\beta$ related to magnetization for parent SrRuO$_3$ closely matches with that for mean field model, and the value of $\beta$ increases with $x$. In following, we have undertaken detailed critical analysis to understand the effect of site dilution on critical behavior in SrRu$_{1-x}$Ti$_x$O$_3$ series.
    
\begin{figure}
	\centering
		\includegraphics[width=8cm]{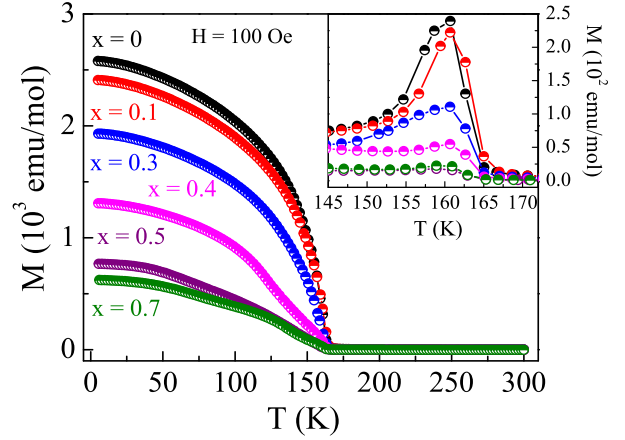}
	\caption{Temperature dependent field cooled magnetization data measured in presence of 100 Oe applied field have been shown for SrRu$_{1-x}$Ti$_x$O$_3$ series. The inset shows zero field cooled magnetization (100 Oe) in expanded temperature range close to $T_c$.}
	\label{fig:Fig1}
\end{figure}

\subsection{Arrott Plot}
The Arrott plot offers an important tool to study the second-order magnetic phase transition and the critical behavior across second order PM-FM phase transition.\cite{arrott1} Arrott plot is about plotting of isothermal $M(H)$ data in form of $M^2$ vs $H/M$ where for mean-field model with $\beta$ = 0.5 and $\gamma$ = 1, the isotherms form a set of parallel straight lines. In another sense, straight line behavior in high field regime in Arrott plot implies magnetic interaction is of mean-field type. Intercept due to straight line fitting in Arrott plot on $M^2$ and $H/M$ axis directly gives spontaneous magnetization ($M_s$) and initial inverse susceptibility ($\chi_0^{-1}$, respectively. Moreover, isotherm which passes through origin in Arrott plot marks the $T_c$ as implies zero $M_s$. As discussed, for analysis of Arrott plot, a set of isotherms $M(H)$ is required across the $T_c$. Figs. 2a - 2f show isotherms ($M$ vs $H$ plots) collected at different temperatures with temperature interval of $\Delta T$ = 1 K across $T_c$ for SrRu$_{1-x}$Ti$_x$O$_3$ series. The $M(H)$ plots for all the samples look like FM type where it shows downward curvature. Inset in each figure of Fig. 2 shows a derivative of magnetization (d$M$/d$H$) as a function of field for one representative temperature. Inset figures show a decreasing slope of $M(H)$ plot with field. This constitutes a typical signature of second-order PM to FM transition and justifies for following critical analysis. 

\begin{figure}[th]
	\centering
		\includegraphics[width=8cm]{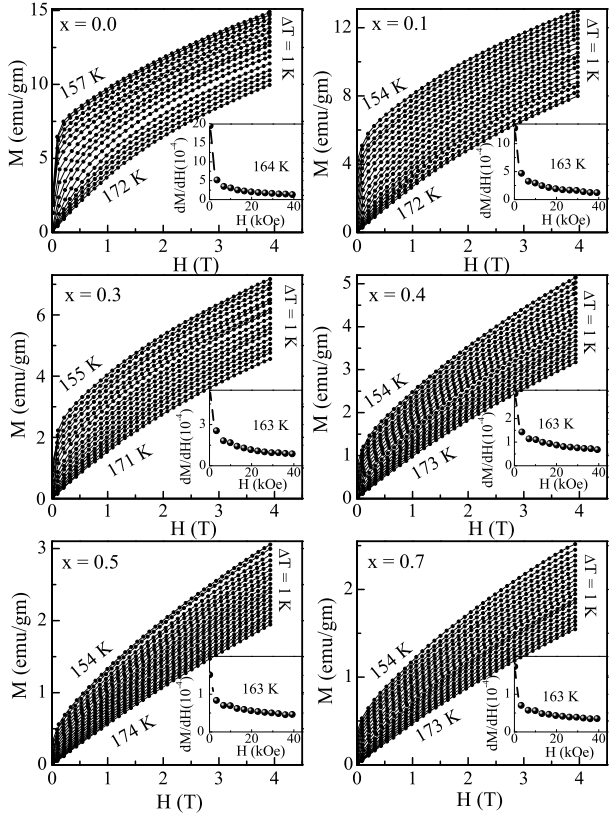}
	\caption{Field dependent isothermal magnetization data near to the transition temperature ${\sim T_c}$ are shown for SrRu$_{1-x}$Ti$_x$O$_3$ series with (a) $x$ = 0.0 (b) $x$ = 0.1 (c) $x$ = 0.3 (d) $x$ = 0.4 (e) $x$ = 0.5 and (f) $x$ = 0.7 composition. Inset of each figure shows magnetic-field derivative of magnetization (d$M$/d$H$) as a function of magnetic field for $M(H)$ plots taken at $T_c$ for respective sample.}
	\label{fig:Fig2}
\end{figure}

\begin{figure}[th]
	\centering
		\includegraphics[width=8cm]{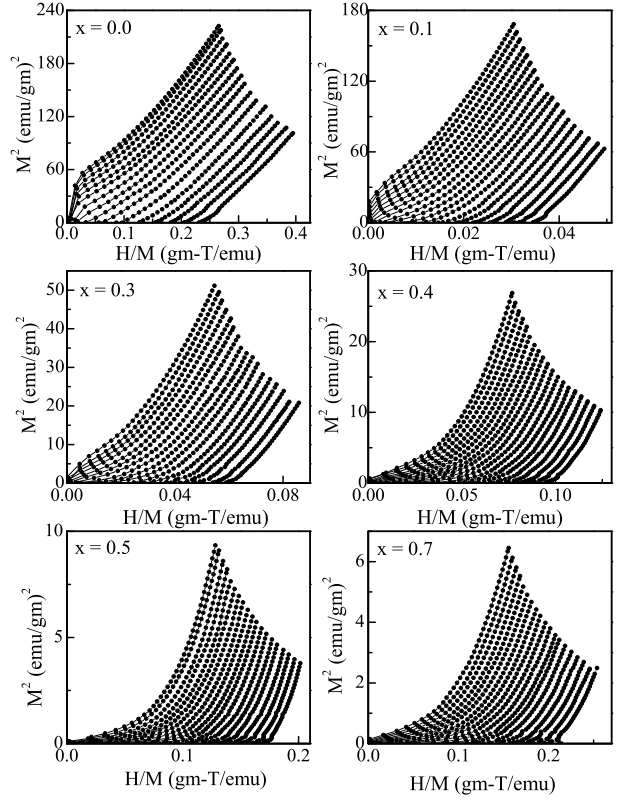}
	\caption{Isotherms in Fig. 2 are plotted in form of Arrott plot ($M^2$ vs $H/M$) for SrRu$_{1-x}$Ti$_x$O$_3$ series with (a) $x$ = 0.0 (b) $x$ = 0.1 (c) $x$ = 0.3 (d) $x$ = 0.4 (e) $x$ = 0.5 and (f) $x$ = 0.7.}
	\label{fig:Fig3}
\end{figure}

Figs. 3a - 3f show Arrott plot, constructed from $M(H)$ isotherm data shown in Fig. 2 for SrRu$_{1-x}$Ti$_x$O$_3$ series. For parent SrRuO$_3$, nature of magnetic state has mostly been shown by earlier studies to follow mean-field interaction model.\cite{fuchs} As discussed, for mean-field model with critical exponents $\beta$ = 0.5 and $\gamma$ = 1, the Arrott plot should yield set of parallel straight lines. Arrott plot for SrRuO$_3$ in Fig. 3a show isotherms do not form straight lines even in high field regime, they are rather slightly curved in upward direction. This means critical exponents for SrRuO$_3$ do not exactly match with mean-field model but they are close to the mean-field values. Fig. 3 further shows with increasing $x$, nonlinearity in Arrott plot increases which suggests nature of magnetic interaction moves away from mean-field model as Ti is introduced in SrRuO$_3$. As we do not obtain parallel straight lines with mean-field exponents, the analysis in Fig. 3 suggests new set of exponents need to be identified for straight line behavior.

To determine the critical exponents and temperature correctly for the present series of samples, we have employed modified Arrott plot (MAP) which is generalized form of Arrott plot and is based on Arrott-Noakes equation of state given as,\cite{arrott2}

\begin{align}
	\left(\frac{H}{M}\right)^{1/\gamma} = a\frac{T-T_c}{T} + b M^{1/\beta}
\end{align}
  
where $a$ and $b$ are the constant.	It is obvious that for $\beta$ = 0.5 and $\gamma$ = 1, Eq. 5 recovers Arrott plot discussed above. Following MAP, isotherms are plotted in form of $M^{1/\beta}$ vs $(H/M)^{1/\gamma}$. Here, we note that we have also tried to form MAP with the exponents $\beta$ and $\gamma$ which are theoretically predicted for 3-dimensional models such as, 3D Heisenberg, 3D Ising, 3D XY models, etc. (Table I) but remain unsuccessful in getting parallel straight lines. As the exponents $\beta$ = 0.5 and $\gamma$ = 1 do not yield straight lines in Fig. 3, we have tuned the exponents $\beta$ and $\gamma$ to obtain parallel straight lines. However, tuning of these parameters is not a straightforward job as two unknown parameters are involved and this often leads to erroneous results. Here, we have adopted an iterative method where we initially tuned $\beta$ and $\gamma$ in Eq. 5 to get apparently good parallel straight lines in modified Arrott plot.\cite{ashim,imtiaz} Then, temperature dependent $M_s(T)$ and $\chi_0^{-1}(T)$ have been obtained from intercept of straight fitting in MAP on $M^{1/\beta}$ and $(H/M)^{1/\gamma}$ axis, respectively. The $T_c$ has been identified as the temperature whose $M(H)$ isotherm passes through origin in MAP. These $M_s(T)$, $\chi_0^{-1}(T)$and $T_c$ are then used in Eq. 1 and 2 to obtain new set of $\beta$ and $\gamma$. These new values of $\beta$ and $\gamma$ have been used to construct similar MAP. This process has been continued till a convergence in values of critical exponents $\beta$ and $\gamma$ and critical temperature $T_c$ is obtained. Figs. 4a - 4f show the MAP of isotherms shown in Figs. 2a - 2f in vicinity of $T_c$ for all the samples. For parent SrRuO$_3$, we find that $\beta$ = 0.54 and $\gamma$ = 0.75 yield parallel straight lines where isotherm at 164 K passes thorough origin marking the $T_c$ of this material (Fig. 4a). The same iterative method has been followed for all the samples. Figs. 4a - 4f show the MAP for all the samples in present series where reasonably good parallel straight lines have been formed in high field regime with proper choice of exponents $\beta$ and $\gamma$. 
  
\begin{figure}[t]
	\centering
		\includegraphics[width=8cm]{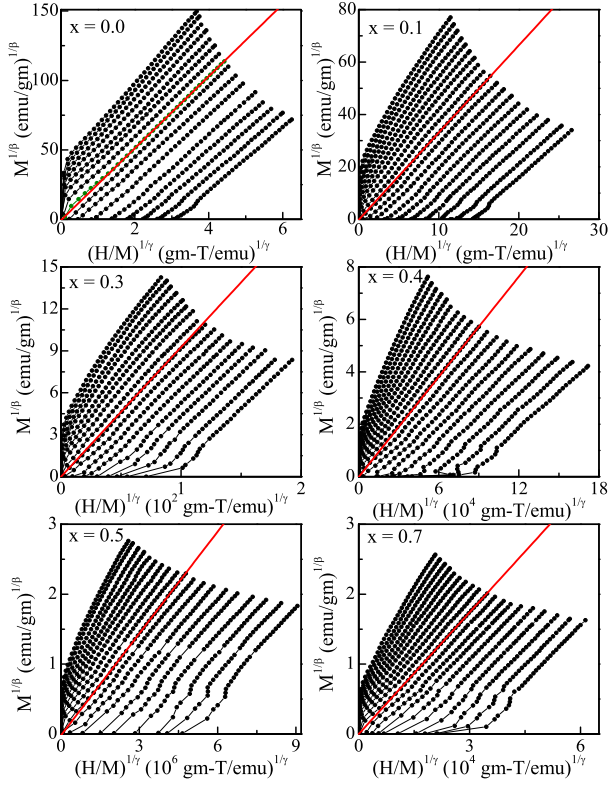}
	\caption{Modified Arrott plot, $M^{1/\beta}$ vs $(H/M)^{1/\gamma}$, constructed out of $M(H)$ data in Fig. 2 are shown for SrRu$_{1-x}$Ti$_x$O$_3$ series with (a) $x$ = 0.0 (b) $x$ = 0.1 (c) $x$ = 0.3 (d) $x$ = 0.4 (e) $x$ = 0.5 and (f) $x$ = 0.7. The solid red line in each figure is due to straight line fitting of modified Arrott plot related to isotherm which passes through origin.}
	\label{fig:Fig4}
\end{figure} 	

\begin{figure}[t]
	\centering
		\includegraphics[width=8cm]{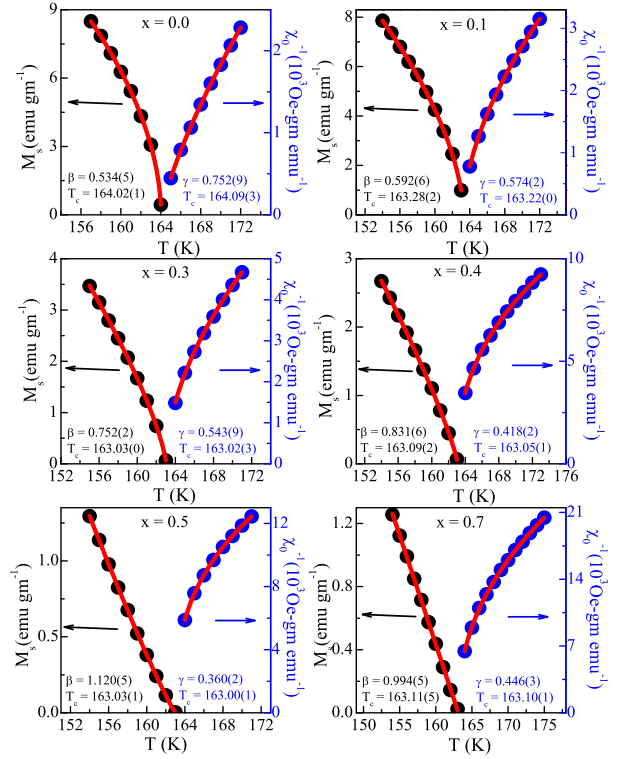}
	\caption{Temperature dependent spontaneous magnetization M$_s$ (left axis) and initial inverse susceptibility $\chi_0^{-1}(T)$ (right axis) as estimated from modified Arrott plots in Fig. 4 are shown for SrRu$_{1-x}$Ti$_x$O$_3$ series with (a) $x$ = 0.0 (b) $x$ = 0.1 (c) $x$ = 0.3 (d) $x$ = 0.4 (e) $x$ = 0.5 and (f) $x$ = 0.7. The solid red lines are due to fitting with power law behavior as stated in Eq. 1 and 2.}
	\label{fig:Fig5}
\end{figure}

The finally obtained $M_s$ and $\chi_0^{-1}$ have been shown as a function of temperature in Figs. 5a - 5f, showing $M_s$ cease to exist once temperature approaches $T_c$. The $M_s(T)$ and $\chi_0(T)$ data in Figs. 5a - 5f have been fitted with Eqs. 1 and 2, respectively for all the compositions. The exponents $\beta$ and $\gamma$ and temperature $T_c$ are obtained as fitting parameters. Table I shows the exponents $\beta$ and $\gamma$ and respective $T_c$ following MAP method. From fitting of $M_s(T)$ with Eq. 1 we obtain $\beta$ = 0.534(5) and $T_c$ = 164.02(1) K and from fitting of $\chi_0^{-1}(T)$ with Eq. 2 we get $\gamma$ = 0.752(9) and $T_c$ = 164.09(3) K. These values of $\beta$, $\gamma$ and $T_c$ are significantly close to the values obtained using MAP in Fig. 4 which underlines the correctness of our iterative method. The $T_c$ values obtained from both $M_s(T)$ and $\chi_0(T)$ in MAP show consistent behavior for whole series. For SrRuO$_3$, obtained $T_c$ $\sim$ 164 K matches well with the reported values. Remarkably, $T_c$ remains almost constant with Ti substitution in present series which substantiates our earlier results. \cite{gupta} The obtained exponents $\beta$ and $\gamma$ do not exactly match with the values theoretically predicted for three dimensional systems (Table I), however, values look closer to mean-field model for $x$ = 0.0 material.         

\subsection{Kouvel-Fisher plot}
Alternatively, critical exponents $\beta$ and $\gamma$ and the critical temperature $T_c$ have been determined using more accurate Kouvel-Fisher (KF) method where the $M_s$ and $\chi_0^{-1}$ are analyzed using Eqs. 6 and 7, respectively.\cite{kouvel} The KF plot for whole series of samples has been shown in Figs. 6a - 6f where $M_s(T)$[d$M_s(T)$/d$T$)]$^{-1}$ vs $T$ and $\chi_0^{-1}(T)$[d$\chi_0^{-1}(T)$/d$T$]$^{-1}$ vs $T$ are plotted. According to KF method (Eq. 6 and 7), these plots would result in straight line behavior where the respective slope would give 1/$\beta$ and 1/$\gamma$. In addition, $T_c$ can be obtained directly and independently as the intercept on temperature axis would give $T_c$. In deed, Fig. 6 shows KF plot forms reasonable straight lines for all the samples which confirms the correctness of KF analysis. For SrRuO$_3$, we obtain exponents $\beta$ = 0.542(8) and $T_c$ = 164.02(3), and $\gamma$ = 0.753(6) and $T_c$ = 164.08(3). These values of $\beta$, $\gamma$ and $T_c$ obtained from KF plot are given in Table I for all the compositions. It is remarkable that values of exponents and $T_c$ obtained from KF plots match closely with those obtained from MAF analysis. With Ti substitution, while $\beta$ increases away from the mean-field value and, on other hand, the $\gamma$ decreases. Nonetheless, agreement between values obtained from two independent MAP and KF method is quite remarkable indicating obtained values are correct.  

\begin{eqnarray}
	 M_s(T)[dM_s(T)/dT]^{-1} = \frac{T - T_c}{\beta}
\end{eqnarray}

\begin{eqnarray}
	 \chi_0^{-1}(T)[d\chi_0^{-1}(T)/dT]^{-1} = \frac{T - T_c}{\gamma}
\end{eqnarray}  

\begin{figure}[th]
	\centering
		\includegraphics[width=8cm]{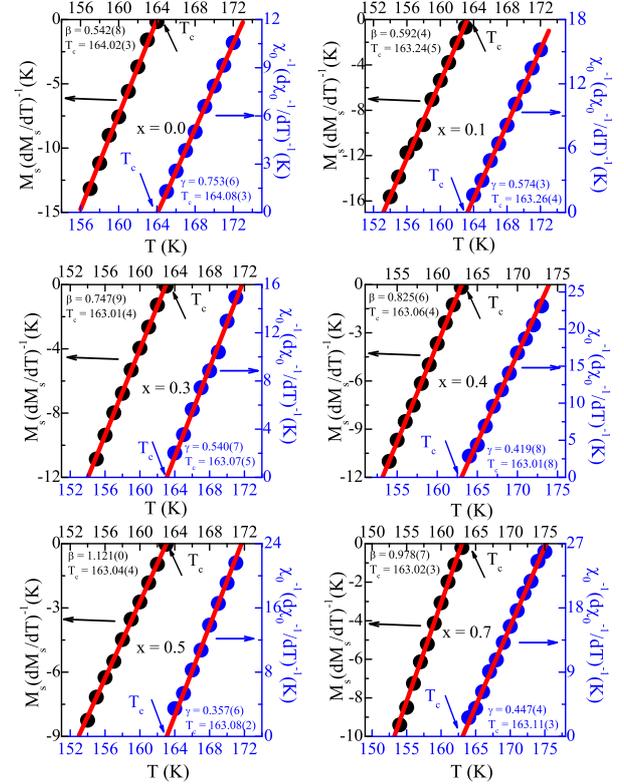}
	\caption{Kouvel-Fisher plots for spontaneous magnetization (Eq. 6) on left axis and for initial inverse susceptibility (Eq. 7) on right have been shown for SrRu$_{1-x}$Ti$_x$O$_3$ series with (a) $x$ = 0.0 (b) $x$ = 0.1 (c) $x$ = 0.3 (d) $x$ = 0.4 (e) $x$ = 0.5 and (f) $x$ = 0.7. The solid straight lines are the linear fits following Eq. 6 and 7.}
	\label{fig:Fig6}
\end{figure}

\subsection{Critical isotherm  plot}
Scaling analysis predicts that variation of $M(H)$ at $T_c$ (critical isotherm) follows a power-law behavior involving exponent $\delta$ (Eq. 3). Critical isotherms are already identified from MAP in Fig. 4. The main panel of Figs. 7a - 7f shows critical isotherms $M(H,T_c)$ for all samples in present series. The inset of Figs. 7a - 7f show same plot in $\log_{10}$-$\log_{10}$ scale. Following Eq. 3, slope due to straight line fitting in $\log M$ vs $\log H$ plot would give 1/$\delta$. The inset plots show reasonably linear behavior, and the exponent $\delta$ calculated from slope is given in Table I as critical isotherm method. Again, obtain $\delta$ does not match with the theoretical values predicted for 3D based different universality classes as mentioned in Table I. To crosscheck the consistency of the estimated exponents, we have used Widom scaling relation where the critical exponents $\beta$, $\gamma$ and $\delta$ obey following relation,\cite{widom}

\begin{eqnarray}
	\delta = 1 + \frac{\gamma}{\beta}
\end{eqnarray}

Eq. 8 indirectly gives exponent $\delta$ using the values of exponents $\gamma$ and $\beta$. The $\delta$ has been calculated following Eq. 8 using $\beta$ and $\gamma$ values which are estimated from both MAP and KF plot methods. Table I shows exponent $\delta$ where the values represent both obtained from analysis of critical isotherm as well as calculated from using Eq. 8. For SrRuO$_3$, we obtain $\delta$ = 2.389(3) following critical isotherm analysis (Fig. 7) and 2.408(2) and 2.389(2) from MAP and KF plot using Eq. 8. Both the estimated and calculated values match well. With Ti substitution, $\delta$ continuously decreases till $x$ = 0.5 and then shows an increased value for $x$ = 0.7. It is remarkable that values of $\delta$ obtained from two distinct process agree reasonably for all the materials which emphasizes that values of exponents $\beta$, $\gamma$, $\delta$ and temperature $T_c$ are estimated quite accurately.  
   
\begin{figure}[th]
	\centering
		\includegraphics[width=8cm]{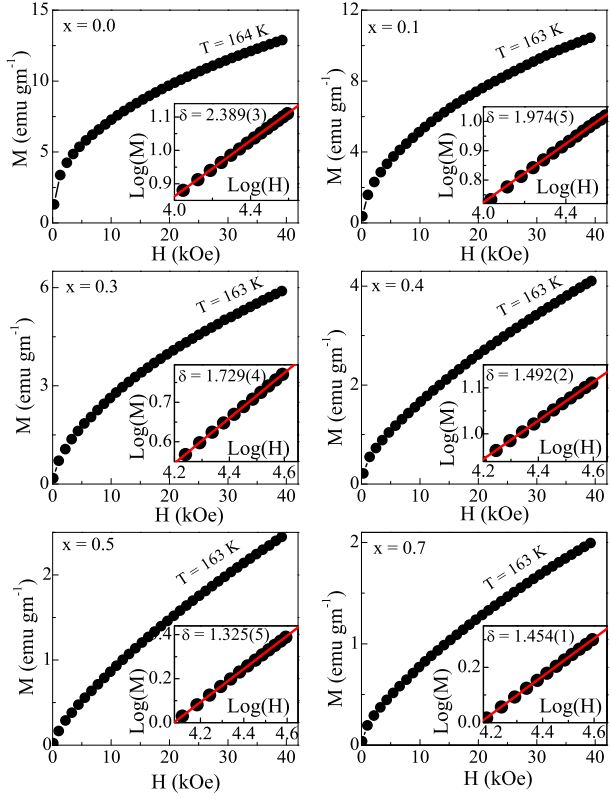}
	\caption{The $M(H)$ isotherm at $T_c$ has been shown for SrRu$_{1-x}$Ti$_x$O$_3$ series with (a) $x$ = 0.0 (b) $x$ = 0.1 (c) $x$ = 0.3 (d) $x$ = 0.4 (e) $x$ = 0.5 and (f) $x$ = 0.7. The inset shows same $M(H)$ data in $\log_{10}$-$\log_{10}$ scale for respective samples and the solid line is due to linear fit of the data.} 
	\label{fig:Fig7}
\end{figure}

\begin{figure}[th]
	\centering
		\includegraphics[width=8cm]{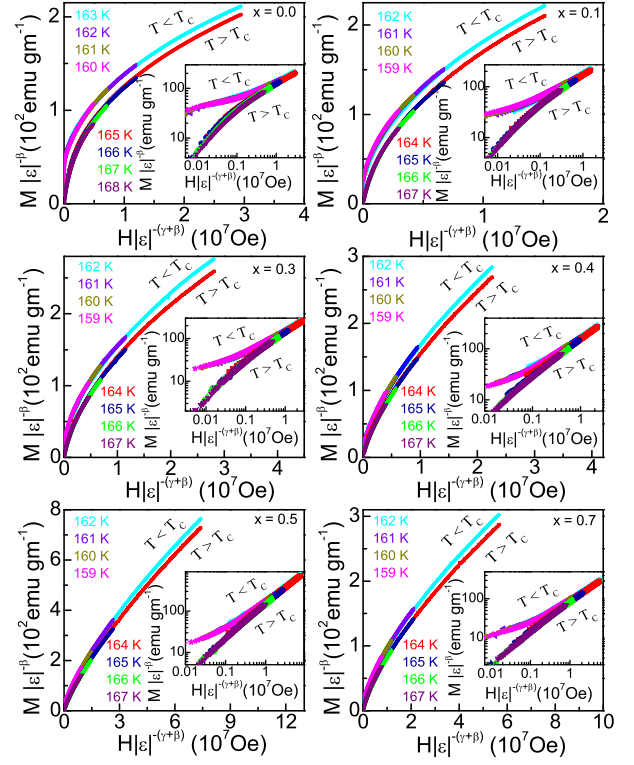}
	\caption{Renormalized magnetization $M\left|t\right|^{-\beta}$ has been plotted as a function of renormalized field $H\left|t\right|^{-(\gamma+\beta)}$ for isotherms both below and above $T_c$. Data are shown for SrRu$_{1-x}$Ti$_x$O$_3$ series with (a) $x$ = 0.0 (b) $x$ = 0.1 (c) $x$ = 0.3 (d) $x$ = 0.4 (e) $x$ = 0.5 and (f) $x$ = 0.7. Inset shows same plot in $\log_{10}$-$\log_{10}$ scale. Renormalized magnetization is merged into single line with renormalized field for isotherms both below and above $T_c$.}
	\label{fig:Fig8}
\end{figure}

\setlength{\tabcolsep}{4pt}
\begin{table*}
\caption{\label{tab:table 1} Table shows the values of critical exponents $\beta$, $\gamma$ and $\delta$ determined using different independent methods such as modifier Arrott plot (MAP), Kouvel-Fisher (KF) plot and critical isotherm analysis for SrRu$_{1-x}$Ti$_x$O$_3$ series. The values of exponents calculated theoretically following different spin interaction models are also given for comparison.}
\begin{tabular}{cccccccc}
\hline
Composition &Ref. &Method &$\beta$ &$T_c(M_s)$ &$\gamma$ &$T_c(\chi_0)$ &$\delta$\\
\hline
SrRuO$_3$ &This work &MAP &0.534(5) &164.02(1) &0.752(9) &164.09(3) &2.408(2)$\footnotemark[1]$\\
 & &KF Method &0.542(8) &164.02(3) &0.753(6) &164.08(3) &2.389(2)$\footnotemark[1]$\\  
 & &Critical Isotherm & & & & &2.389(3)\\
\hline
SrRu$_{0.9}$Ti$_{0.1}$O$_3$ &This work &MAP &0.592(6) &163.28(6) &0.574(2) &163.22(0) &1.969(5)$\footnotemark[1]$\\
 & &KF Method &0.592(4) &163.24(5) &0.574(3) &163.26(4) &1.969(5)$\footnotemark[1]$\\ 
 & &Critical Isotherm & & & & &1.974(5)\\
\hline
SrRu$_{0.7}$Ti$_{0.3}$O$_3$ &This work &MAP &0.752(2) &163.03(2) &0.543(9) &163.02(3) &1.722(0)$\footnotemark[1]$\\
 & &KF Method &0.747(9) &163.01(4) &0.540(7) &163.07(3) &1.722(8)$\footnotemark[1]$\\  
 & &Critical Isotherm & & & & &1.729(4)\\
\hline
SrRu$_{0.6}$Ti$_{0.4}$O$_3$ &This work &MAP &0.831(6) &163.09(2) &0.418(2) &163.05(1) &1.503(0)$\footnotemark[1]$\\
 & &KF Method &0.825(6) &163.06(4) &0.419(8) &163.01(8) &1.507(8)$\footnotemark[1]$\\  
 & &Critical Isotherm & & & & &1.492(2)\\
\hline
SrRu$_{0.5}$Ti$_{0.5}$O$_3$ &This work &MAP &1.120(5) &163.03(1) &0.360(2) &163.00(1) &1.321(4)$\footnotemark[1]$\\
 & &KF Method &1.121(0) &163.04(4) &0.357(6) &163.08(2) &1.318(4)$\footnotemark[1]$\\  
 & &Critical Isotherm & & & & &1.325(5)\\
\hline
SrRu$_{0.3}$Ti$_{0.7}$O$_3$ &This work &MAP &0.994(5) &163.11(5) &0.446(3) &163.10(1) &1.448(6)$\footnotemark[1]$\\
 & &KF Method &0.978(7) &163.02(3) &0.447(4) &163.11(3) &1.457(0)$\footnotemark[1]$\\  
 & &Critical Isotherm & & & & &1.454(1)\\
\hline
Mean-field Model  &\cite{kaul} &Theory &0.5 & &1.0 & &3.0\\
\hline
3D Heisenberg Model  &\cite{kaul} &Theory &0.365 & &1.386 & &4.8\\
\hline
3D Ising Model  &\cite{kaul} &Theory &0.325 & &1.241 & &4.82\\
\hline
\end{tabular}
\end{table*} 
\footnotetext[1]{Calculated following Eq. 8}

\subsection{Scaling analysis}
So far we have seen that critical exponents in present series do not exactly match with any established theoretical models for classical spin interaction. However, the exponents and $T_c$ determined using different independent methods such as, modified Arrott plot, Kouvel-Fisher plot and critical isotherm analysis agree very closely (Table I). Moreover, exponents follow Widom relation. We have further tested the consistency and accuracy of exponents as well as $T_c$ using a scaling analysis (Eq. 4). This analysis suggests plotting of magnetic isotherm $M(H)$ data in form of $M\left|t\right|^{-\beta}$ vs $H\left|t\right|^{-(\gamma + \beta)}$ where the isotherms both below and above $T_c$ are scaled into a separate line for correct values of criitical exponents and $T_c$. This rather constitutes a rigorous test to examine the consistency and accuracy of exponents as well as $T_c$. Scaling plots are shown in Figs. 8a - 8f for this present series where four scaled isotherms both below and above of respective $T_c$ are plotted following Eq. 4. As seen in Fig. 8, isotherms both below and above $T_c$ are nicely scaled into a line . The same plot are shown in $\log_{10}$-$\log_{10}$ scale in inset of Fig. 8 for better presentation which also shows a nice scaling of isotherm data. This reconfirms that estimated exponents ($\beta$, $\gamma$ and $\delta$) as well as $T_c$ in Table I for all the samples in this series are accurate and consistent.

\subsection{Evolution of exponents in SrRu$_{1-x}$Ti$_x$O$_3$}
Our experimental data analysis show critical exponent $\beta$ for SrRuO$_3$ is close to the mean-field value 0.5, however, exponents $\gamma$ and $\delta$ show the values lower than those for mean-field model i.e., 1.0 and 3.0, respectively (see Table I). For SrRuO$_3$, we obtain $T_c$ $\sim$ 164 K which is very consistent with reported values for this material,\cite{allen,cao,fuchs,cheng,chanchal} even this $T_c$ is one of the highest value obtained for this material. Our critical analysis interestingly show that substitution of Ti has very little influence on $T_c$ of SrRuO$_3$ as we find $T_c$ remains almost unchanged. This is in conformity with our earlier study where we have explained this constant behavior of $T_c$ in SrRu$_{1-x}$Ti$_x$O$_3$ considering opposite change of electron correlation $U$ and density of states N($\epsilon_F$) that maintains $T_c$ in picture of itinerant ferromagnet.

Nonetheless, critical exponents show an unusual evolution where $\beta$ increases and both $\gamma$ and $\delta$ decreases with $x$ (see Table I). In Fig. 9, we have shown an evolution of exponents $\beta$, $\gamma$ and $\delta$ with Ti ($x$) normalized by respective values for SuRuO$_3$ ($x$ = 0). The figure shows that $\beta$ continuously increases and $\gamma$, $\delta$ decreases till $x$ = 0.5, though for $x$ = 0.7 the values appear to show a reverse turn which may be related to metal-to-insulator transition around $x$ $\sim$ 0.4 in SrRu$_{1-x}$Ti$_x$O$_3$ series.\cite{kim-MI} Here, we mention that critical exponents and $T_c$ are obtained using different independent techniques namely, MAF, KF plot and critical isotherm analysis and they nicely agree. It can be noted that determined exponents and $T_c$ fairly obey the scaling law behavior (Fig. 8) for all the samples. Moreover, the Widom relation (Eq. 8), which demonstrates relation among the critical exponents is nicely followed for entire range of Ti doping (Table I). These attest to the fact that determined exponents and $T_c$ are consistent and correct with experimental accuracy. These exponents show a steady variation and deviates from mean-field behavior with increasing $x$. However, the exponent values do not match with the predicted values for different known universality class models, also those can not be explained with theoretical models available at hands for spin exchange interaction. Note, that similar kind of evolution of critical exponents ($\beta$, $\gamma$ and $\delta$) has been evidenced in Sr$_{1-x}$Ca$_x$RuO$_3$, however, prominent difference is that $T_c$ decreases to zero around 70\% of Ca doping resulting in quantum phase transition whereas $T_c$ remains almost unchanged in present series.\cite{fuchs,cheng,gupta} In case of Sr$_{1-x}$Ca$_x$RuO$_3$, this particular evolution of exponents has been attributed to crossover from mean-field like behavior to disorder-induced strong coupling regime arising from quantum fluctuations near quantum phase transition point by Fuchs \textit{et al.}\cite{fuchs} An another study reports that following of Arrott plot (signature of mean-field model) in SrRuO$_3$ may not be associated with itinerant model of magnetism, rather this arises due to continuous curvature evolution from Sr$_{0.5}$Ca$_{0.5}$RuO$_3$ to SrRuO$_3$ to BaRuO$_3$ as induced by changes in lattice distortion and spin-orbit coupling effect.\cite{cheng} These changes are responsible for band narrowing in Sr$_{1-x}$Ca$_x$RuO$_3$ which leads to phase segregation between strongly and weakly correlated phases, hence exponents evolve accordingly. Here, it can be further noted that in heavy fermion compound URu$_{1-x}$Re$_x$Si$_2$, while though exponent $\beta$ remains constant but the exponents $\gamma$ and ($\delta$ - 1) decreases to zero as the QPT is approached with decreasing Re substitution around 15\%.\cite{butch}

While though the magnetic moment and the Curie-Weiss temperature decreases in SrRu$_{1-x}$Ti$_x$O$_3$ but the ferromagnetic $T_c$ does not modify unlike in case of Sr$_{1-x}$Ca$_x$RuO$_3$.\cite{fuchs,jin,cheng} Moreover, minimum structural modification due to almost matching ionic size between Ru$^{4+}$ and Ti$^{4+}$ in present SrRu$_{1-x}$Ti$_x$O$_3$ series would unlikely cause this peculiar changes in exponents. This system, however, does not convert into first order type PM-FM transition with Ti as seen from decreasing slope in $M(H)$ plot in Fig. 2. Nonetheless, this unusual change in exponents ($\beta$, $\gamma$ and $\delta$) is quite intriguing as their values substantially depart from mean-field values ($\beta$ = 0.5, $\gamma$ = 1.0 and $\delta$ = 3.0), and also from the values typically realized for classical ferromagnetic systems ($\beta$ $<$ 0.5, $\gamma$ $>$ 1.0 and $\delta$ $>$ 3.0). We believe that Ti substituted at Ru-site dilutes the magnetic exchange path, and at higher doping concentration of Ti the system develops into small-size magnetic clusters which is manifested through Griffiths phase like behavior, as reported in our earlier study.\cite{gupta} This has another striking similarity with Sr$_{1-x}$Ca$_x$RuO$_3$ which also exhibits GP like behavior with progressive doping of Ca.\cite{jin,cheng} The disorder is prerequisite for the GP property which is provided by structural distortion in Sr$_{1-x}$Ca$_x$RuO$_3$ whereas site dilution achieved by Ti substitution acts as source of disorder in present SrRu$_{1-x}$Ti$_x$O$_3$. Henceforth, we speculate that effect of disorder realized from both Ti as well as magnetic clusters can provide possible explanation for this unusual change in critical exponents where $\beta$ increases and both $\gamma$ and $\delta$ decreases. The value of $\gamma$ lower than unity in SrRuO$_3$ suggests parent $x$ = 0 sample has some disorder or phase inhomogeneity above $T_c$. This effect of these changes in exponents is quite evident in Fig. 1 where the PM-FM phase transition broadens with progressive Ti substitution. In fact, a recent optical spectroscopy study has shown the temperature dependence of optical conductivity spectra indicates itinerant-type FM in SrRuO$_3$.\cite{jeong} This study further demonstrates non-vanishing local spin moment (sub-micron size) at $T$ $>$ $T_c$ (PM state) originating from local-band exchange splitting which is explained with fluctuating local band theory. However, the net moment is realized to be zero due to spatial and temporal fluctuations of local bands. This needs to be understood whether this renders a constant $T_c$ in present SrRu$_{1-x}$Ti$_x$O$_3$ series.

\begin{figure}
	\centering
		\includegraphics[width=8.5cm]{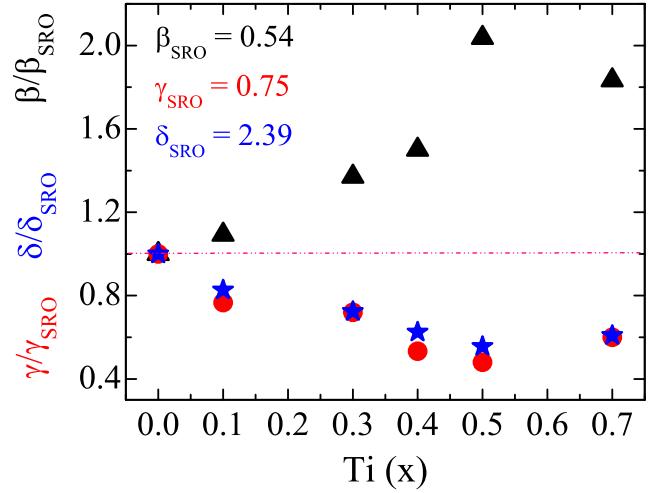}
	\caption{The critical exponents scaled by its values for $x$ = 0.0 compound are shown as a function of composition ($x$) for SrRu$_{1-x}$Ti$_x$O$_3$ series.}
	\label{fig:Fig9}
\end{figure} 

\section{Conclusion}
In summary, we have studied the critical behavior in perovskite based SrRu$_{1-x}$Ti$_x$O$_3$ as a function of Ti substitution using the standard methods such as, modified Arrott plot, Kouvel-Fisher plot and critical isotherm analysis. We have estimated critical exponents $\beta$, $\gamma$ and $\delta$ where $\beta$ increases and both $\gamma$ and $\delta$ decreases with Ti substitution. The transition temperature $T_c$, however, remains almost unchanged with site dilution by Ti doping. The estimated exponents do not match with the values of any universality classes known for spin interaction models. Nonetheless, the exponents nicely obey the Widom relation as well as scaling behavior which attest to the fact that estimated exponents and $T_c$ are consistent and accurate. The evolution of exponents of similar nature has been observed in isoelectronic doped Sr$_{1-x}$Ca$_x$RuO$_3$. This specific change of exponents is likely caused by disorders arising from magnetic clusters which originates due to site dilution with Ti substitution. The evolution of Griffiths phase like behavior in present series as evidenced in our earlier study substantiates the formation of magnetic clusters and its influences on the magnetic behavior.

\section{Acknowledgment}   
We acknowledge Advanced Instrumentation Research Facility (AIRF), JNU for magnetization measurements. We thank Mr. Saroj Jha for the help during the measurements. RG and INB acknowledge the financial support from UGC, India for BSR fellowship and from CSIR, India for SRF Fellowship, respectively.

\end{document}